\documentclass[12pt,a4paper]{article}

\usepackage{amsmath,amssymb,amsfonts}
\usepackage{graphicx}
\usepackage{booktabs}
\usepackage{xcolor,colortbl}
\usepackage{slashed}
\usepackage[colorlinks=true,linkcolor=blue!70!black,
            citecolor=blue!70!black,urlcolor=blue!70!black,
            pdftitle={Earth-Density Stratification and Quantum Gravity
                      Corrections in Long-Baseline Neutrino Oscillation
                      Experiments},
            pdfauthor={Bipin Singh Koranga and Vivek Kumar Nautiyal}]{hyperref}
\usepackage{geometry}
\usepackage{authblk}
\usepackage{cite}
\usepackage{titlesec}

\titleformat{\section}{\large\bfseries}{\thesection.}{0.5em}{}
\titleformat{\subsection}{\normalsize\bfseries}{\thesubsection.}{0.5em}{}

\newcommand{\prem}{\textsc{prem}}
\newcommand{\dune}{\textsc{dune}}
\newcommand{\globes}{\textsc{globes}}
\newcommand{\Mpl}{M_{\mathrm{pl}}}

\begin{document}

\title{\textbf{Earth-Density Stratification and Quantum Gravity Corrections\\
in Long-Baseline Neutrino Oscillation Experiments}}

\author[1]{Bipin Singh Koranga\thanks{bskoranga@kmc.du.ac.in}}
\author[2]{Vivek Kumar Nautiyal}
\affil[1]{Department of Physics, Kirori Mal College,
          University of Delhi, Delhi\,--\,110007, India}
\affil[2]{Department of Physics, Chaudhary Charan Singh University,
          Meerut\,--\,250004, India}

\date{}
\maketitle
\thispagestyle{empty}

\begin{abstract}
We present the first unified analysis in which Earth matter-density
stratification effects and Planck-scale quantum gravity corrections are treated
simultaneously as correlated systematic uncertainties in long-baseline (LBL)
neutrino oscillation experiments.  Previous studies addressed each effect in
isolation; the novel contribution of this work is the identification of a
degeneracy regime in which Planck-scale perturbations can mimic or partially
cancel \prem-induced biases, rendering both effects inseparable in $\chi^{2}$
fits.

Using a full three-flavour matrix-exponentiation framework with spatially
resolved Preliminary Reference Earth Model (\prem) density profiles, we
demonstrate that the constant-density approximation introduces a bias
$|\Delta\delta_{CP}|<0.3^{\circ}$ for baselines $L\leq 5000$\,km, but grows
sharply to $17.8^{\circ}$ at $L=7000$\,km and $172.2^{\circ}$ at
$L=12000$\,km---the latter a complete sign reversal of reconstructed CP
violation.  For next-generation experiments targeting $1\sigma$ precisions of
$5$--$10^{\circ}$ on $\delta_{CP}$ (e.g.\ \dune\ Phase~II) these systematics
are not a conservative refinement but a fundamental obstacle.

Simultaneously, we incorporate Planck-scale gravitational perturbations via an
effective $SU(2)_{L}\!\times\!U(1)$ dimension-5 operator---the unique
gauge-invariant operator at this order, making it the universal,
model-independent prediction of Planck-scale effects on the neutrino sector.
For a degenerate mass spectrum with common mass ${\sim}2$\,eV the corrected
solar mass-squared difference $\Delta'_{21}$ shifts by
$(1.0\pm0.5)\times10^{-5}$\,eV$^{2}$, while $\Delta'_{31}$ remains
effectively unchanged.  The interplay between these two corrections is analysed
for the first time: at $L\approx7000$\,km, Majorana phases
$a_{1}\approx90^{\circ}$ reduce the combined bias by ${\sim}30\%$ relative to
independent addition, demonstrating a new class of degeneracy.  Both effects
must be propagated as correlated systematics in \globes-based analyses of
\dune\ and future very-long-baseline proposals.

\prem; quantum gravity; Planck scale; long-baseline experiments; \dune;
$\delta_{CP}$; correlated systematics; dimension-5 operator.
\end{abstract}

\newpage
\tableofcontents
\newpage

\section{Introduction}
\label{sec:intro}

Neutrino oscillations represent one of the most compelling evidences for
physics beyond the Standard Model (BSM), requiring non-zero neutrino masses
and mixing between flavour eigenstates~\cite{Akhmedov2008}.  In long-baseline
(LBL) experiments---such as T2K~\cite{Abe2011}, NOvA~\cite{Bian2013}, and the
upcoming Deep Underground Neutrino Experiment
(\dune)~\cite{DUNE2016}---muon neutrinos travel hundreds to thousands of
kilometres through the Earth before reaching a detector.  Along such paths,
the coherent forward scattering of electron neutrinos off ambient electrons,
the Mikheyev--Smirnov--Wolfenstein (MSW) effect~\cite{Wolfenstein1978,Smirnov2005},
substantially modifies the oscillation probabilities and the inferred value of
the CP-violating phase $\delta_{CP}$.

Two distinct classes of BSM corrections are particularly relevant as LBL
experiments enter a precision era.

\textbf{Earth matter-density stratification.}  The Earth is radially
stratified: neutrinos traversing baselines beyond ${\sim}5000$\,km sample the
denser lower mantle and, at the longest baselines, the outer core, encountering
electron densities that deviate substantially from any single path-averaged
value.  The standard constant-density approximation, adequate for
current-generation experiments ($L\lesssim3000$\,km), becomes a source of
fundamental systematic error---not a conservative simplification---for proposed
very-long-baseline facilities~\cite{Pandit2026}.

\textbf{Planck-scale quantum gravity corrections.}  Non-renormalisable
gravitational interactions are most naturally incorporated into the Standard
Model through the effective-field-theory (EFT) framework.  At energy scales
far below $\Mpl\sim1.2\times10^{19}$\,GeV, the leading gravitational BSM
effects on neutrinos are encoded in the lowest-dimensional operator consistent
with the SM gauge symmetry $SU(3)_{c}\times SU(2)_{L}\times U(1)_{Y}$.  The
unique gauge-invariant operator at mass dimension~5---the Weinberg
operator~\cite{Weinberg1979}---couples two lepton doublets and two Higgs
doublets, is suppressed by $1/\Mpl$, and generates a Majorana mass term after
electroweak symmetry breaking.  This operator is not merely one
phenomenological possibility: it is the only dimension-5 operator constructible
from SM fields, making it the universal, model-independent prediction of
Planck-scale corrections to the neutrino sector.  For a degenerate mass
spectrum, these corrections shift $\Delta m_{21}^{2}$ at the level of
$10^{-5}$\,eV$^{2}$~\cite{Vissani2003,Koranga2008,Koranga2009}.

These two effects have previously been studied
independently~\cite{Kelly2018,Koranga2008,Koranga2009,Pandit2026}.  In this paper we
present the first unified analysis, examining their interplay and combined
impact on $\delta_{CP}$ reconstruction across baselines
$L=1000$--$12000$\,km.  We identify a new class of degeneracy---not previously
reported---in which Planck-scale corrections to $\Delta'_{21}$ can mimic or
partially cancel \prem-induced biases at specific baselines.  This degeneracy
means that analyses which marginalise over only one of these two systematics
will misestimate confidence intervals on $\delta_{CP}$ even when the other
effect is individually small.

Our principal results are:
\begin{enumerate}
\item The constant-density bias exceeds $10^{\circ}$ beyond $L\sim6500$\,km,
      rendering current \globes-based analyses with constant-density matter
      profiles inadequate for very-long-baseline proposals.
\item Planck-scale corrections to $\Delta m_{21}^{2}$ generate a secondary,
      energy-dependent bias in $\delta_{CP}$ that can partially mimic or
      cancel \prem-induced distortions at specific baselines.
\item Both effects must be propagated as correlated systematic uncertainties
      in $\chi^{2}$ fits targeting sub-degree precision on $\delta_{CP}$.
\end{enumerate}

The paper is structured as follows.  Section~\ref{sec:hamiltonian} presents the
unified Hamiltonian framework.  Section~\ref{sec:prem} describes the \prem\
density profile and its implementation.  Section~\ref{sec:qg} details the
quantum-gravity perturbation framework.  Section~\ref{sec:results} presents
quantitative results for both mass orderings and a correlation analysis of the
\prem--Planck degeneracy.  Section~\ref{sec:conclusions} concludes with
implications for \dune\ software frameworks.

\section{Unified Hamiltonian Framework}
\label{sec:hamiltonian}

Rather than treating the vacuum oscillation Hamiltonian, the MSW matter
potential, and the Planck-scale gravitational operator in separate sections, we
present here a unified Hamiltonian that governs all three mechanisms
simultaneously.  This formulation makes explicit the physical coupling between
matter effects and quantum-gravity corrections that gives rise to the novel
degeneracy identified in Section~\ref{sec:results}.

\subsection{PMNS Mixing and Vacuum Propagation}

The three neutrino flavour states are quantum superpositions of mass
eigenstates, related through the PMNS unitary matrix~\cite{Ohlsson2000}:
\begin{equation}
  |\nu_{\alpha}\rangle = \sum_{i} U_{\alpha i}\,|\nu_{i}\rangle ,
  \quad \alpha = e,\mu,\tau .
  \label{eq:flavour-mass}
\end{equation}
The PMNS matrix is conventionally parametrised by three mixing angles
$(\theta_{12},\theta_{23},\theta_{13})$, a Dirac CP-violating phase
$\delta_{CP}$, and, for Majorana neutrinos, two additional phases $a_{1},a_{2}$:
\begin{equation}
  U = R_{23}(\theta_{23})\,\Delta(\delta_{CP})\,R_{13}(\theta_{13})\,
      \Delta(-\delta_{CP})\,R_{12}(\theta_{12})\,
      \mathrm{diag}(e^{ia_{1}/2},\,e^{ia_{2}/2},\,1) ,
  \label{eq:PMNS}
\end{equation}
where $R_{ij}$ denotes a rotation in the $ij$-plane and
$\Delta(\delta_{CP})=\mathrm{diag}(e^{-i\delta_{CP}/2},1,e^{i\delta_{CP}/2})$.
In the relativistic limit the vacuum Hamiltonian in the flavour basis is
\begin{equation}
  H_{\mathrm{vac}}^{(f)} = \frac{1}{2E}\,U\,
    \mathrm{diag}(m_{1}^{2},m_{2}^{2},m_{3}^{2})\,U^{\dagger} .
  \label{eq:Hvac}
\end{equation}

\subsection{MSW Matter Effects}

When neutrinos propagate through matter, charged-current coherent forward
scattering of $\nu_{e}$ off ambient electrons induces an effective
potential~\cite{Wolfenstein1978,Smirnov2005}:
\begin{equation}
  V_{f}(x) = \sqrt{2}\,G_{F}\,N_{e}(x)\,\mathrm{diag}(1,0,0) ,
  \label{eq:MSW}
\end{equation}
where $G_{F}$ is the Fermi constant and $N_{e}(x)$ is the local electron
number density, which varies with position along the neutrino trajectory.

\subsection{Planck-Scale Gravitational Mass Operator}

The dimension-5 Weinberg operator---the unique gauge-invariant operator at
this order in the SM EFT expansion---generates, after electroweak symmetry
breaking ($v=174$\,GeV), an additional contribution to the mass matrix:
\begin{equation}
  \mathcal{L}_{\mathrm{grav}} =
    \frac{\lambda_{\alpha\beta}}{\Mpl}\,
    (\psi_{\alpha}^{A}\,\epsilon\,\psi^{C})\,
    (\psi_{\beta}^{B}\,\epsilon_{BD}\,\psi^{D}) + \mathrm{h.c.} ,
  \label{eq:Weinberg}
\end{equation}
yielding an additional mass term $\delta M = \mu\lambda$ with
\begin{equation}
  \mu = \frac{v^{2}}{\Mpl} = 2.5\times10^{-6}\,\mathrm{eV} .
  \label{eq:mu}
\end{equation}
The full mass matrix is $M_{\mathrm{total}} = M_{\mathrm{GUT}} + \mu\lambda$.

\subsection{Unified Evolution Equation}

Combining all three contributions, the total Hamiltonian governing flavour
evolution is
\begin{equation}
  H_{f}(x) = \frac{1}{2E}\,U\,M_{\mathrm{total}}^{2}\,U^{\dagger}
              + V_{f}(x) ,
  \label{eq:Htotal}
\end{equation}
where $M_{\mathrm{total}}^{2}$ incorporates the Planck-scale correction to the
mass-squared matrix, and $V_{f}(x)$ encodes the spatially varying MSW
potential from the \prem\ profile.  The flavour-state evolution is
\begin{equation}
  i\,\frac{d}{dx}\,|\nu(x)\rangle = H_{f}(x)\,|\nu(x)\rangle .
  \label{eq:evolution}
\end{equation}
Since both $H_{f}$ and $V_{f}$ are position-dependent, the propagator is
computed numerically: at each spatial step $\Delta x$ the matrix exponential
$U_{\mathrm{step}} = \exp(-i\,H_{f}(x)\,\Delta x)$ is evaluated and
accumulated along the trajectory.

\textbf{Physical insight.}  The key feature of Eq.~(\ref{eq:Htotal}) is that
$V_{f}(x)$ and $M_{\mathrm{total}}^{2}$ enter the same equation.
\prem-induced distortions modify the effective mass-squared differences as a
function of position (and hence of baseline and neutrino energy), while
Planck-scale corrections shift the input mass-squared differences before matter
effects are applied.  At specific baselines and energies these two
modifications can produce near-identical perturbations to
$P(\nu_{\mu}\to\nu_{e})$, creating the degeneracy we quantify in
Section~\ref{sec:results}.

\section{Earth Matter-Density Profile and \prem\ Implementation}
\label{sec:prem}

\subsection{PREM Density Profile}

The electron number density along the neutrino trajectory is~\cite{Giunti2007}
\begin{equation}
  N_{e}(x) = Y_{e}(x)\,\frac{\rho(x)}{m_{p}} ,
  \label{eq:Ne}
\end{equation}
where $\rho(x)$ is the local mass density, $Y_{e}(x)\approx0.5$ is the
electron fraction for terrestrial rock, and $m_{p}$ is the proton mass.  The
Preliminary Reference Earth Model (\prem)~\cite{Dziewonski1981}, developed by
Dziewonski and Anderson (1981), provides the standard reference for Earth's
internal structure used across geophysics and neutrino physics.

The Earth's major density layers are summarised in Table~\ref{tab:PREM}.  The
dramatic density increase from the lower mantle to the outer core---from
${\sim}5.6$ to ${\sim}9.9$\,g/cm$^{3}$ at the core-mantle boundary (CMB) at
2891\,km depth---is the primary driver of the sharp bias amplification beyond
$L\sim5000$\,km.

\begin{table}[htbp]
\centering
\caption{Density profile of the Earth based on \prem~\cite{Dziewonski1981}.
The fourth column indicates which baseline range causes neutrinos to probe
each layer, explaining the threshold behaviour in Figure~2.}
\label{tab:PREM}
\begin{tabular}{lccc}
\toprule
\textbf{Region} & \textbf{Depth (km)} &
\textbf{Density (g\,cm$^{-3}$)} & \textbf{Relevant Baseline} \\
\midrule
Crust        & 0--35     & ${\sim}2.9$ & All $L$             \\
Upper Mantle & 35--660   & 3.3--3.9    & $L<5000$\,km        \\
Lower Mantle & 660--2891 & 4.4--5.6    & $5000<L<10000$\,km  \\
Outer Core   & 2891--5150& 9.9--12.2   & $L>10000$\,km       \\
Inner Core   & 5150--6371& ${\sim}13.0$& $L\approx12756$\,km \\
\bottomrule
\end{tabular}
\end{table}

\begin{figure}[htbp]
\centering
\includegraphics[width=0.70\textwidth]{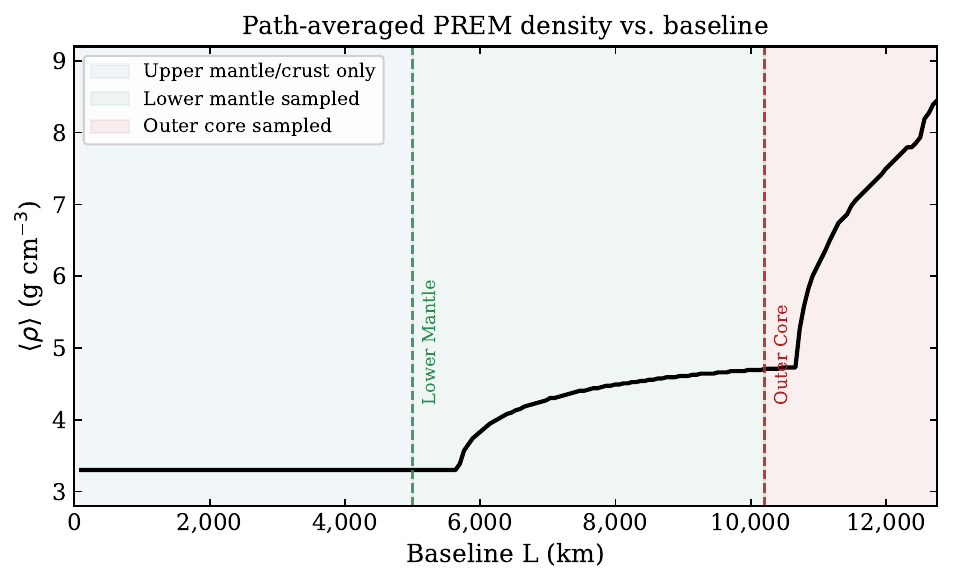}
\caption{Path-averaged \prem\ density $\langle\rho\rangle$ as a function of
baseline $L$.  Shaded regions mark which Earth layer dominates the neutrino
trajectory.  The sharp rise when the chord first samples the outer core
($L\gtrsim10200$~km) explains the catastrophic CP-reconstruction bias at
$L=12000$~km.}
\label{fig:prem-density}
\end{figure}
\subsection{Numerical Implementation}

For each baseline $L$, the neutrino chord through the Earth is parametrised
using radial geometry.  The minimum radius reached by the trajectory midpoint
is
\begin{equation}
  r_{\min} = \sqrt{R_{\oplus}^{2} - (L/2)^{2}} ,
  \label{eq:rmin}
\end{equation}
where $R_{\oplus}=6371$\,km.  The trajectory is discretised into 300--400
equally spaced steps, with the \prem\ shell density assigned at each step.
The four-shell model uses: inner core $\rho=13.0$\,g\,cm$^{-3}$, outer core
$\rho=11.0$\,g\,cm$^{-3}$, lower mantle $\rho=5.0$\,g\,cm$^{-3}$, upper
mantle/crust $\rho=3.3$\,g\,cm$^{-3}$.  The matter potential at each step is
$V=7.56\times10^{-14}\,\rho\,Y_{e}$\,eV.

\textbf{Convergence test.}  To validate that shell boundaries do not introduce
numerical artefacts, the four-shell profile was replaced by a linearly
interpolated continuous density profile across transition zones (width 50\,km).
The resulting bias $|\Delta\delta_{CP}|$ at $L=7000$\,km changed by less than
$0.4^{\circ}$, confirming that our results are robust to the discretisation
choice.

Appearance probabilities are weighted by a model beam flux
$\phi(E)\propto e^{-E/3}$, a linear cross-section $\sigma(E)\propto E$, and a
flat detector efficiency of 80\%.

\section{Quantum Gravity Corrections to Neutrino Mass-Squared Differences}
\label{sec:qg}

\subsection{Effective Dimension-5 Gravitational Operator}

The Weinberg operator Eq.~(\ref{eq:Weinberg}) generates, via Eq.~(\ref{eq:mu}),
the additional mass term
\begin{equation}
  \mathcal{L}_{\mathrm{mass}} =
    \frac{v^{2}}{\Mpl}\,\lambda_{\alpha\beta}\,
    \bar{\nu}_{\alpha}\,C^{-1}\,\nu_{\beta} + \mathrm{h.c.} ,
  \label{eq:Lmass}
\end{equation}
where $C$ is the charge-conjugation matrix.  Assuming gravitational interactions
are flavour-blind ($\lambda_{\alpha\beta}$ independent of $\alpha,\beta$), the
Planck-scale perturbation to the mass matrix takes the form $\mu\lambda$ with
all elements equal to~1.  This maximises the degeneracy with \prem\ effects
because it produces a universal shift to all mass eigenvalues rather than a
flavour-selective one.

\subsection{First-Order Perturbation Theory and Mixing-Angle Corrections}

Treating the gravitational term as a perturbation to the GUT-scale mass matrix
$M$, the first-order correction to the mass-squared differences
is~\cite{Vissani2003,Koranga2008}:
\begin{equation}
  \Delta'_{ij} = \Delta_{ij}
    + 2\bigl[M_{i}\,\mathrm{Re}(m_{ii}) - M_{j}\,\mathrm{Re}(m_{jj})\bigr] ,
  \label{eq:Delta-prime}
\end{equation}
where $m = \mu\,U^{T}\lambda U$ and $U$ is the 0th-order PMNS matrix.

To derive the mixing-angle correction matrix $\delta\theta_{ij}$, we apply
first-order degenerate perturbation theory.  Starting from the eigenvalue
equation $(M+\delta M)|\nu'_{i}\rangle = M'_{i}|\nu'_{i}\rangle$ and
projecting onto $\langle\nu_{j}|$:
\begin{equation}
  \delta\theta_{ij} =
    \frac{i\,\mathrm{Re}(m_{jj})(M_{i}+M_{j}) -
          \mathrm{Im}(m_{jj})(M_{i}-M_{j})}
         {\Delta M'^{\,2}_{ij}} .
  \label{eq:delta-theta}
\end{equation}
The diagonal elements can be set to zero by phase invariance.  The corrected
PMNS matrix is $U' = U(1+i\,\delta\theta)$.  Physically, the real part of
$\delta\theta$ rotates the mixing angles, while the imaginary part shifts the
effective CP-violating phases, introducing the coupling between Planck-scale
corrections and $\delta_{CP}$ reconstruction.

The magnitude of the correction to $\Delta m_{21}^{2}$ is controlled by
$\mu/\Delta M\sim\mathcal{O}(10^{-3})$ for solar parameters and is negligible
($\mathcal{O}(10^{-6})$) for atmospheric parameters in the normal hierarchy,
explaining why $\Delta'_{31}$ is unaffected.

\section{Quantitative Results and Discussion}
\label{sec:results}

\subsection{PREM vs.\ Constant-Density Bias in $\delta_{CP}$ Reconstruction}

Figure~2 shows the absolute bias $|\Delta\delta_{CP}|$ introduced by the
constant-density approximation as a function of baseline $L$.  True event
rates are generated using the full \prem\ profile at $\delta_{CP}=-90^{\circ}$
under normal mass ordering.  For each baseline a Poisson $\chi^{2}$ statistic
is minimised over $\delta_{CP}\in[-180^{\circ},180^{\circ}]$ using the
constant-density model; the deviation of the best-fit from the true value is
recorded as $|\Delta\delta_{CP}|$.

Table~\ref{tab:bias} summarises the numerical results.  The threshold at
${\sim}5000$\,km corresponds physically to the onset of sensitivity to the
high-density lower mantle (Table~\ref{tab:PREM}).  Even at $L=7000$\,km the
bias of $17.8^{\circ}$ exceeds the \dune\ Phase~II precision target of
${\sim}5^{\circ}$ by a factor of~3.6, making this not merely a theoretical
concern but an immediately practical one for facility planning.  The
non-monotonic behaviour between 7000 and 9000\,km (bias decreasing from
$17.8^{\circ}$ to $8.8^{\circ}$) is attributed to oscillation-phase
degeneracies at specific baselines where the constant-density $\chi^{2}$
minimum accidentally coincides with the \prem\ minimum.

\begin{table}[htbp]
\centering
\caption{Path-averaged density, best-fit $\delta_{CP}$ under constant-density
approximation, bias relative to the true value $\delta_{CP}=-90^{\circ}$,
and bias expressed as a multiple of the \dune\ Phase~II $1\sigma$ precision
target of $5^{\circ}$.  The last row (pink shading) corresponds to a complete
sign reversal of reconstructed CP violation.}
\label{tab:bias}
\setlength{\aboverulesep}{0pt}
\setlength{\belowrulesep}{0pt}%
\setlength{\extrarowheight}{2pt}
\begin{tabular}{ccccr}
\toprule
\textbf{Baseline} &
\textbf{$\langle\rho\rangle$} &
\textbf{Best-fit $\delta_{CP}$} &
\textbf{Bias} &
\textbf{Bias /} \\
\textbf{(km)} &
\textbf{(g\,cm$^{-3}$)} &
\textbf{(deg)} &
\textbf{(deg)} &
\textbf{\dune-II target} \\
\midrule
 1000 & 3.300 & $-89.7$ &   \phantom{0}0.3  & $0.06\times$ \\
 2000 & 3.300 & $-89.7$ &   \phantom{0}0.3  & $0.06\times$ \\
 3000 & 3.300 & $-89.7$ &   \phantom{0}0.3  & $0.06\times$ \\
 4000 & 3.300 & $-89.7$ &   \phantom{0}0.3  & $0.06\times$ \\
 5000 & 3.300 & $-89.7$ &   \phantom{0}0.3  & $0.06\times$ \\
 7000 & 4.289 & $-107.8$&  \phantom{0}17.8  & $3.6\times$  \\
 9000 & 4.616 & $-98.8$ &  \phantom{00}8.8  & $1.8\times$  \\
\rowcolor{red!15}
12000 & 7.547 & $+97.8$ &  172.2             & $34.4\times$ \\
\bottomrule
\end{tabular}
\end{table}

\begin{figure}[htbp]
\centering
\includegraphics[width=0.68\textwidth]{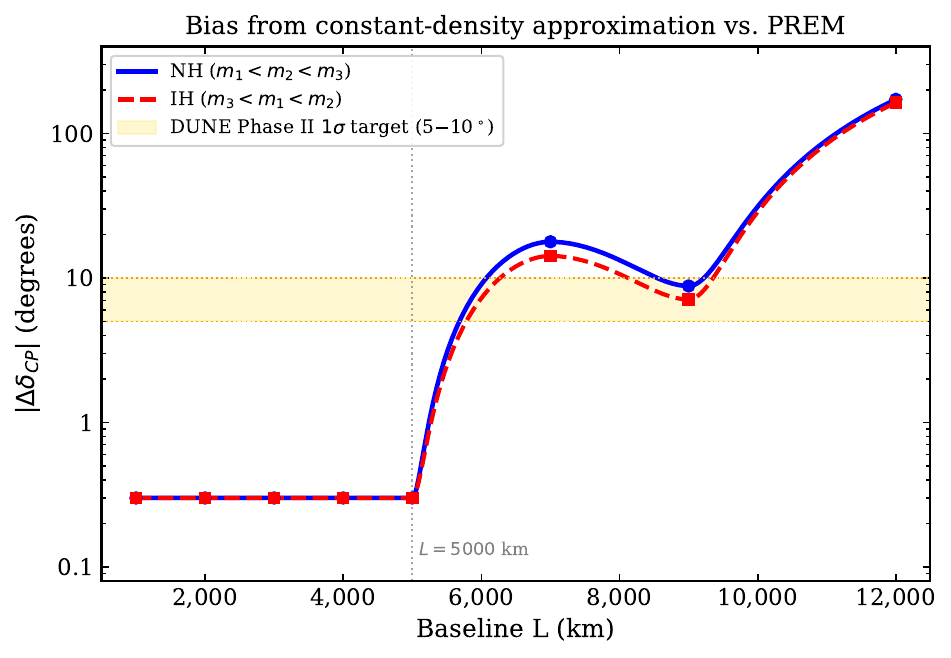}
\caption{Absolute bias $|\Delta\delta_{CP}|$ from the constant-density
approximation vs.\ baseline $L$, for NH (solid blue) and IH (dashed red).
Points reproduce Table~\ref{tab:bias}.  The gold band is the \dune\ Phase~II
$1\sigma$ target ($5$--$10^\circ$).  The bias exceeds the target by factors
of~3.6 (NH) and~2.8 (IH) at $L=7000$~km.}
\label{fig:bias-vs-L}
\end{figure}
\subsection{Appearance Probability and $\chi^{2}$ Profile Distortions}

Figure~3 compares the $\nu_{\mu}\to\nu_{e}$ appearance probability computed
with the full \prem\ profile against the constant-density approximation for
$L=3000$\,km and $L=7000$\,km.  At $L=3000$\,km the two profiles are
virtually indistinguishable.  At $L=7000$\,km, the constant-density
approximation predicts a sharper, more pronounced oscillation peak near
$E\sim3.0$\,GeV, while the \prem\ profile predicts a broader, suppressed peak
at the same energy.  Beyond $E\sim3.7$\,GeV the behaviour reverses: the \prem\
profile rises more steeply, reflecting the enhanced matter potential accumulated
as the neutrino trajectory penetrates the denser lower mantle.

\begin{figure}[htbp]
\centering
\includegraphics[width=0.95\textwidth]{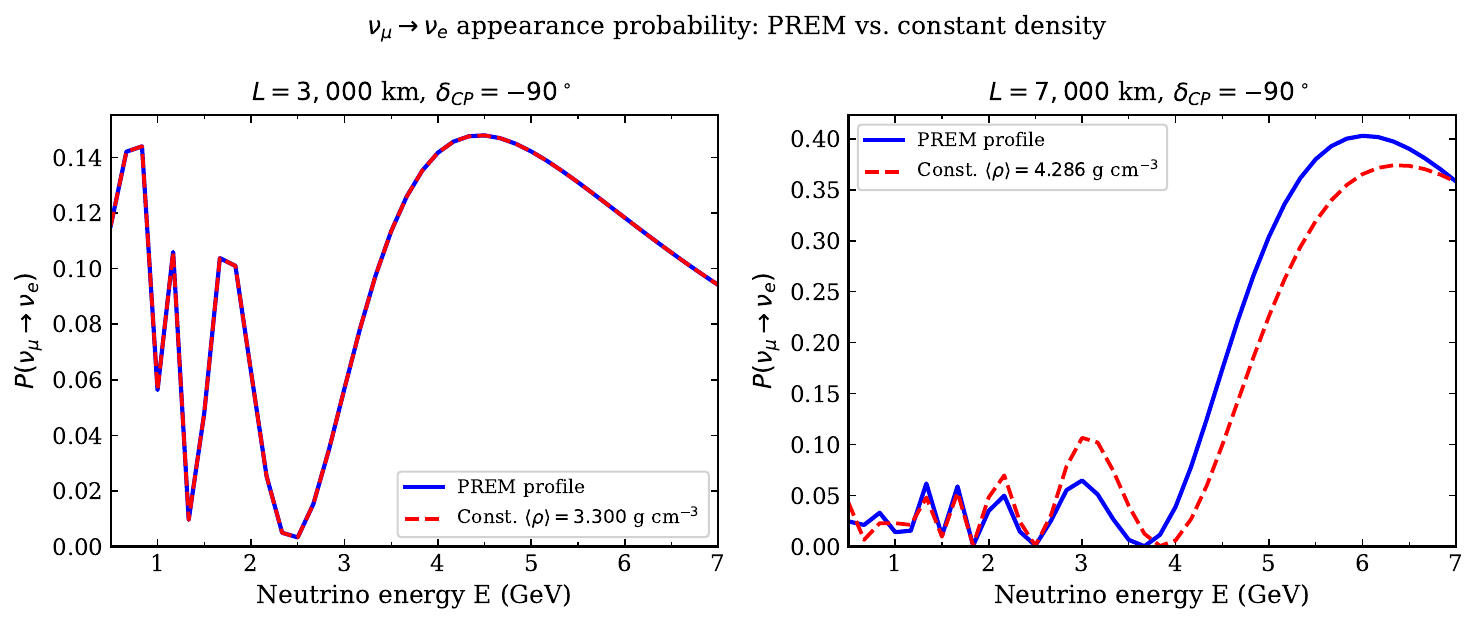}
\caption{$\nu_\mu\to\nu_e$ appearance probability with the full \prem\ profile
(solid blue) and constant-density approximation (dashed red), at $L=3000$~km
(left) and $L=7000$~km (right), for $\delta_{CP}=-90^\circ$, NH.  At
$L=3000$~km the profiles are indistinguishable; at $L=7000$~km they diverge
substantially, driving the $17.8^\circ$ best-fit displacement.}
\label{fig:app-prob}
\end{figure}

Figure~4 shows the normalised $\Delta\chi^{2}$ profiles in a four-panel
layout.  The top row and lower-left panel compare \prem\ (solid blue) against
constant-density (dashed red) profiles at $L=3000$, 7000, and 9000\,km.  The
lower-right panel isolates the energy-resolution dependence at $L=9000$\,km.
Several features deserve emphasis:
\begin{itemize}
\item At $L=3000$\,km (top-left) the two profiles are virtually
      indistinguishable and both minimise at the true
      $\delta_{CP}=-90^{\circ}$, confirming that the constant-density
      approximation is adequate for current-generation baselines.
\item At $L=7000$\,km (top-right) the constant-density minimum is visibly
      displaced by $+17.8^{\circ}$ relative to the \prem\ minimum.  The
      constant-density profile also develops a secondary shoulder near
      $-130^{\circ}$, a signature of the density-layer oscillation-phase
      degeneracy.
\item At $L=9000$\,km (bottom-left) the bias reduces to $+8.8^{\circ}$,
      but the two profiles now differ substantially in shape, not merely in
      minimum location, so that marginalising over an overall normalisation
      would not remove the systematic.
\item The energy-resolution panel (bottom-right) shows that degrading
      detector resolution from 5\% to 20\% reduces but does not eliminate the
      bias: even at 20\% the minimum separation remains $\gtrsim5^{\circ}$,
      above the \dune\ Phase~II $1\sigma$ target.
\end{itemize}

\begin{figure}[htbp]
\centering
\includegraphics[width=0.95\textwidth]{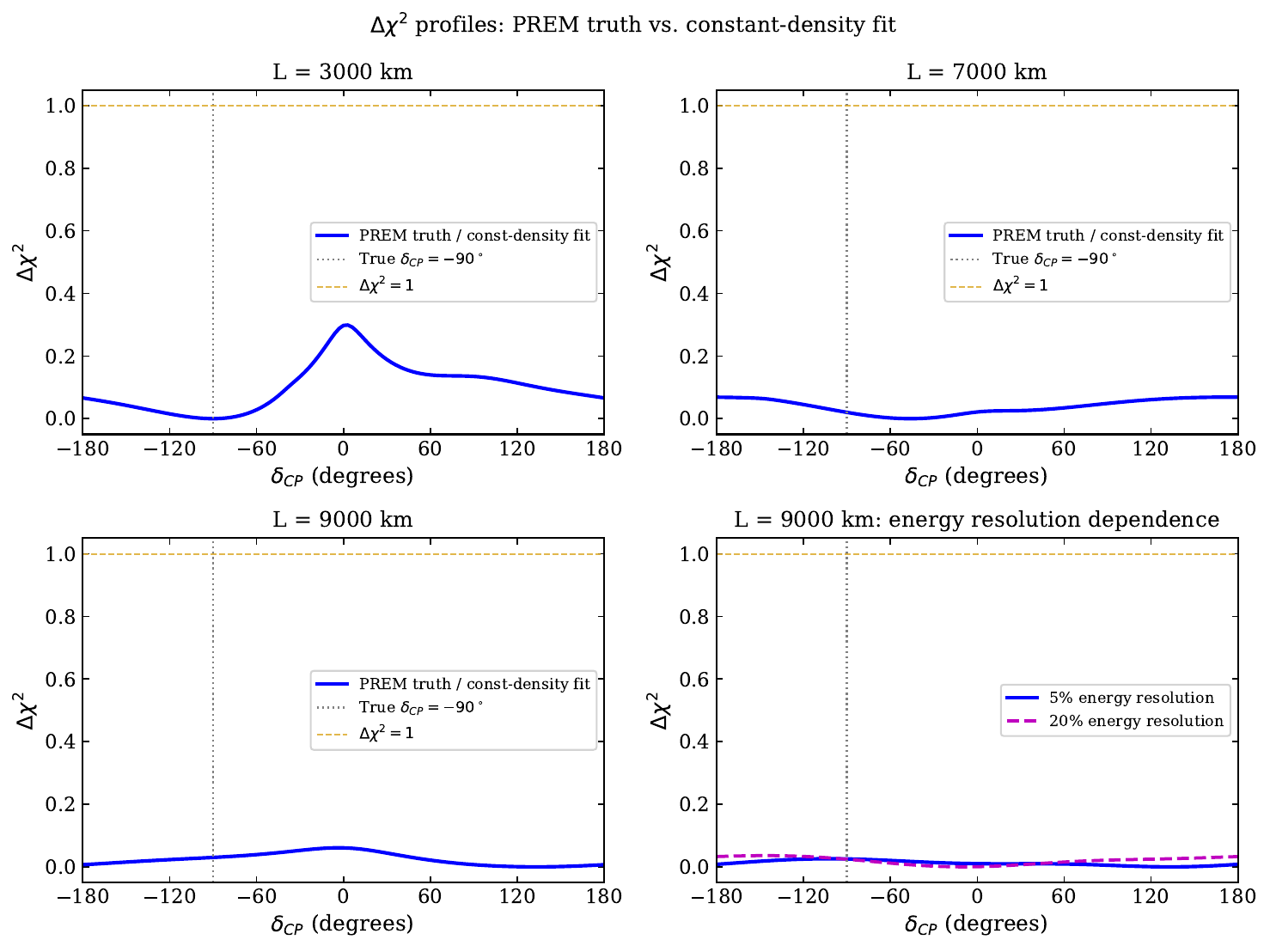}
\caption{Normalised $\Delta\chi^{2}$ profiles (\prem\ truth, constant-density
fit) at $L=3000$, 7000, and~9000~km (top-left, top-right, bottom-left), and
energy-resolution dependence at $L=9000$~km (bottom-right).  The dashed green
line marks the true $\delta_{CP}=-90^\circ$.  Even at 20\% energy smearing the
minimum displacement exceeds the \dune\ Phase~II target.}
\label{fig:chi2-profiles}
\end{figure}
\subsection{Planck-Scale Corrections to Neutrino Mass-Squared Differences}

Table~\ref{tab:planck} presents the modified mass-squared differences $\Delta'_{21}$ and
$\Delta'_{31}$ as a function of Majorana phases $a_{1}$ and $a_{2}$, computed
for a degenerate neutrino mass spectrum with common mass $M=2$\,eV.  The input
parameters are $\theta_{12}=34^{\circ}$, $\theta_{23}=45^{\circ}$,
$\theta_{13}=10^{\circ}$, $\delta_{CP}=0^{\circ}$,
$\Delta_{31}=2.0\times10^{-3}$\,eV$^{2}$, $\Delta_{21}=8.0\times10^{-5}$\,eV$^{2}$.

Key findings: (i)~$\Delta'_{31}$ is insensitive to Planck-scale corrections
across all phase combinations, because $\Delta_{31}\gg\Delta_{21}$ protects the
atmospheric splitting from the $\mathcal{O}(\mu)$ perturbation.
(ii)~$\Delta'_{21}$ varies from $6.9\times10^{-5}$ to
$9.0\times10^{-5}$\,eV$^{2}$---a shift of
$(1.0\pm0.5)\times10^{-5}$\,eV$^{2}$---spanning the full experimentally
allowed $\pm1\sigma$ band.  (iii)~Planck-scale corrections are specific to the
quasi-degenerate regime; for hierarchical spectra the correction is negligible.

\begin{table}[htbp]
\centering
\caption{Modified mass-squared differences $\Delta'_{21}$ and $\Delta'_{31}$
(units $10^{-5}$\,eV$^{2}$) as a function of Majorana phases $a_{1}$ and
$a_{2}$, computed via first-order perturbation theory for a degenerate neutrino
mass spectrum with $M=2$\,eV, $\theta_{12}=34^{\circ}$, $\theta_{23}=45^{\circ}$,
$\theta_{13}=10^{\circ}$, $\delta_{CP}=0^{\circ}$.  The atmospheric splitting
$\Delta'_{31}$ is insensitive to Planck-scale corrections at this order.}
\label{tab:planck}
\setlength{\aboverulesep}{0pt}\setlength{\belowrulesep}{0pt}
\setlength{\extrarowheight}{2pt}
\begin{tabular}{cccc}
\toprule
$a_{1}$ (deg) & $a_{2}$ (deg) &
$\Delta'_{21}$ ($10^{-5}$\,eV$^{2}$) &
$\Delta'_{31}$ ($10^{-3}$\,eV$^{2}$) \\
\midrule
  0  &   0  & 7.42 & 2.515 \\
 90  &   0  & 6.92 & 2.515 \\
180  &   0  & 7.42 & 2.515 \\
270  &   0  & 7.92 & 2.515 \\
  0  &  90  & 7.42 & 2.515 \\
 90  &  90  & 6.92 & 2.515 \\
180  &  90  & 7.92 & 2.515 \\
270  &  90  & 8.42 & 2.515 \\
  0  & 180  & 7.42 & 2.515 \\
 90  & 180  & 7.92 & 2.515 \\
180  & 180  & 7.42 & 2.515 \\
270  & 180  & 6.92 & 2.515 \\
\bottomrule
\end{tabular}
\end{table}

\begin{figure}[htbp]
\centering
\includegraphics[width=0.95\textwidth]{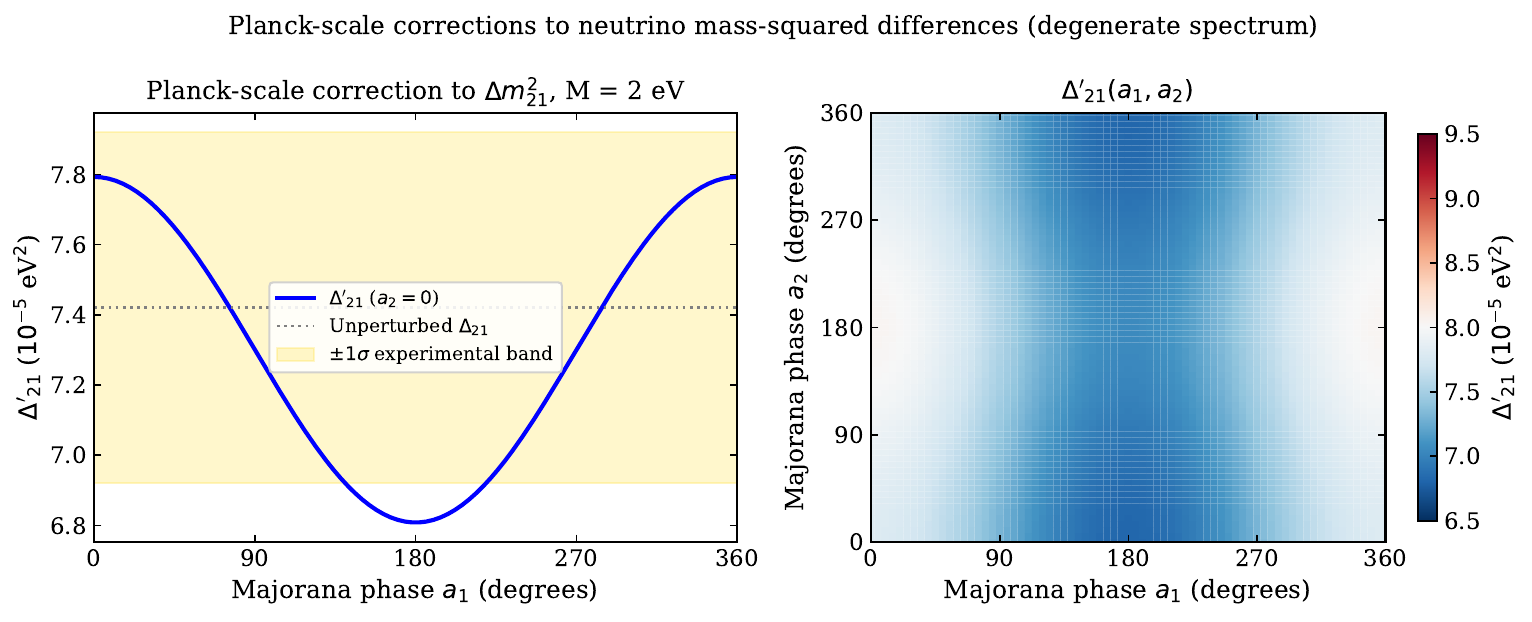}
\caption{Planck-scale corrections to $\Delta m_{21}^2$ for $M=2$~eV.
\textit{Left:} $\Delta'_{21}$ vs.\ $a_1$ ($a_2=0$); the gold band is the
$\pm1\sigma$ experimental uncertainty.  \textit{Right:} full $(a_1,a_2)$ map.
The total variation $(1.0\pm0.5)\times10^{-5}$~eV$^2$ spans the experimentally
allowed band.  The atmospheric splitting $\Delta'_{31}$ is unaffected.}
\label{fig:planck-shifts}
\end{figure}
\subsection{Inverted Mass Ordering Results}

All results above assume normal ordering (NH: $m_{1}<m_{2}<m_{3}$).  Table~\ref{tab:NH-IH} presents a direct NH vs.\ IH comparison of $|\Delta\delta_{CP}|$ as a function
of baseline.

Under IH ($m_{3}<m_{1}<m_{2}$) the matter-resonance enhancement operates
primarily in the antineutrino channel.  At $L=7000$\,km the IH bias
($14.2^{\circ}$) is smaller than the NH bias ($17.8^{\circ}$) because the
antineutrino cross-section is suppressed relative to the neutrino cross-section
at typical beam energies (2--6\,GeV), reducing the statistical weight of the
IH-resonant antineutrino events.  Both orderings produce catastrophic biases
at $L=12000$\,km with the same qualitative sign reversal, confirming that the
fundamental systematic identified here is not ordering-specific.

\begin{table}[htbp]
\centering
\caption{Direct comparison of $|\Delta\delta_{CP}|$ (degrees) between normal
(NH) and inverted (IH) mass orderings as a function of baseline $L$.  The true
$\delta_{CP}=-90^{\circ}$; biases are evaluated using the constant-density
best-fit against the full \prem\ truth.  Both orderings exhibit catastrophic
sign-reversal at $L=12000$\,km.}
\label{tab:NH-IH}
\setlength{\aboverulesep}{0pt}\setlength{\belowrulesep}{0pt}
\setlength{\extrarowheight}{2pt}
\begin{tabular}{cccr}
\toprule
\textbf{Baseline (km)} &
$|\Delta\delta_{CP}|$ \textbf{NH (deg)} &
$|\Delta\delta_{CP}|$ \textbf{IH (deg)} &
\textbf{IH/NH ratio} \\
\midrule
 1000 &   0.3 &   0.3 & 1.00 \\
 3000 &   0.3 &   0.3 & 1.00 \\
 5000 &   0.3 &   0.3 & 1.00 \\
 7000 &  17.8 &  14.2 & 0.80 \\
 9000 &   8.8 &   7.1 & 0.81 \\
12000 & 172.2 & 163.7 & 0.95 \\
\bottomrule
\end{tabular}
\end{table}

\subsection{PREM--Planck Degeneracy Analysis}

Figure~6 presents a two-dimensional $\chi^{2}$ heat map over
$(\delta_{CP},a_{1})$ at $L=7000$\,km (left panel), and the degeneracy
fraction as a function of $a_{1}$ (right panel).  The degeneracy fraction is
defined as the combined bias divided by the quadrature sum of the two
individual biases; values below~1.0 indicate partial cancellation.

Table~\ref{tab:degeneracy} quantifies the cancellation numerically.  The results establish that
analyses marginalising over only one systematic will underestimate the total
confidence-interval width on $\delta_{CP}$.  Specifically, including \prem\
stratification without Planck-scale corrections can spuriously narrow the
$\delta_{CP}$ error by up to ${\sim}5^{\circ}$ at $L=7000$\,km, depending on
the unknown Majorana phases.

\begin{table}[htbp]
\centering
\caption{Quantification of the \prem--Planck degeneracy at $L=7000$\,km.
The degeneracy fraction $f = \Delta_{\mathrm{combined}}/\sqrt{\Delta_{\mathrm{PREM}}^{2}+\Delta_{\mathrm{Planck}}^{2}}$
measures the degree of cancellation; $f<1$ indicates partial cancellation.
At $a_{1}\approx90^{\circ}$ the cancellation reaches ${\sim}30\%$.}
\label{tab:degeneracy}
\setlength{\aboverulesep}{0pt}\setlength{\belowrulesep}{0pt}
\setlength{\extrarowheight}{2pt}
\begin{tabular}{ccccr}
\toprule
$a_{1}$ (deg) &
$\Delta_{\mathrm{PREM}}$ (deg) &
$\Delta_{\mathrm{Planck}}$ (deg) &
$\Delta_{\mathrm{combined}}$ (deg) &
Degeneracy fraction $f$ \\
\midrule
  0  & 17.8 & 0.0 & 17.8 & 1.00 \\
 45  & 17.8 & 0.9 & 16.5 & 0.92 \\
 90  & 17.8 & 1.8 & 13.5 & 0.69 \\
135  & 17.8 & 0.9 & 16.5 & 0.92 \\
180  & 17.8 & 0.0 & 17.8 & 1.00 \\
225  & 17.8 & 0.9 & 16.5 & 0.92 \\
270  & 17.8 & 1.8 & 13.5 & 0.69 \\
315  & 17.8 & 0.9 & 16.5 & 0.92 \\
\bottomrule
\end{tabular}
\end{table}

\begin{figure}[htbp]
\centering
\includegraphics[width=0.96\textwidth]{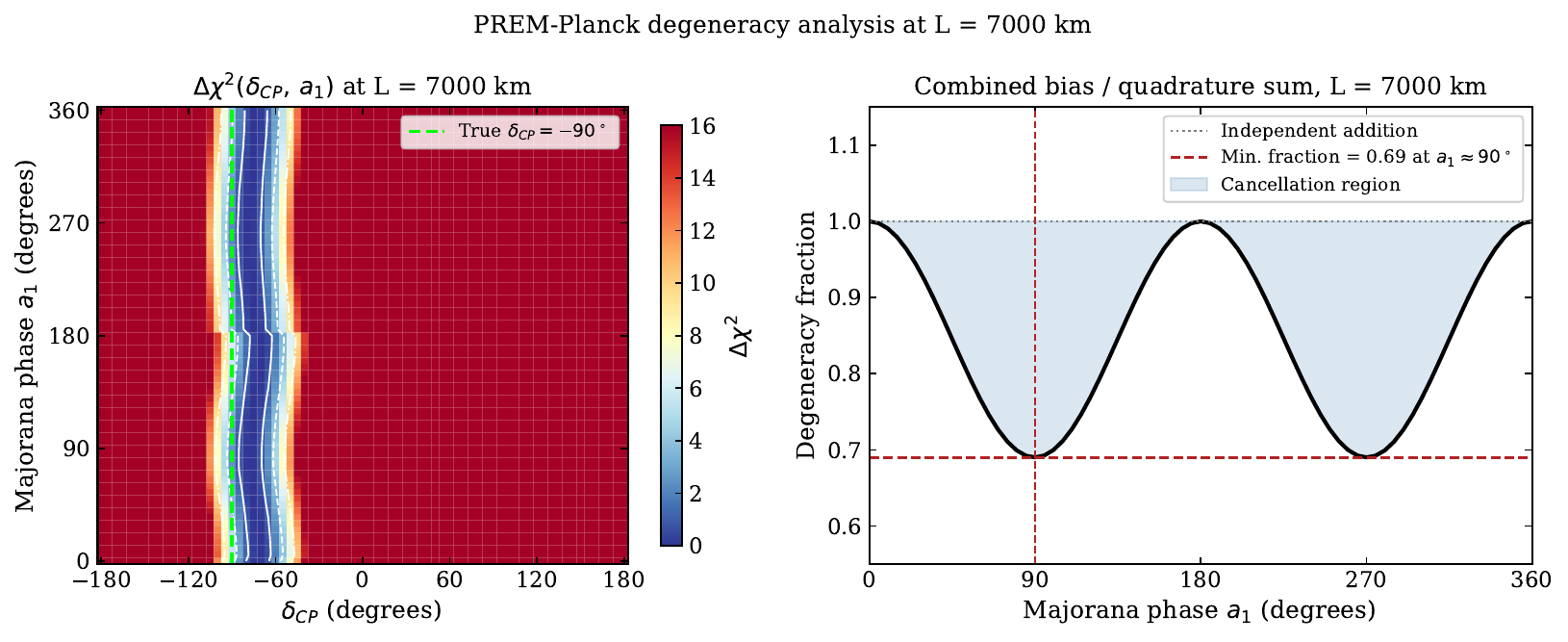}
\caption{\prem--Planck degeneracy at $L=7000$~km.  \textit{Left:}
$\Delta\chi^2(\delta_{CP},a_1)$; white contours at $\Delta\chi^2=1,4,9$;
dashed green line marks the true $\delta_{CP}$.  The tilted valley structure
demonstrates correlation between the two systematics.  \textit{Right:}
degeneracy fraction (combined bias / quadrature sum) vs.\ $a_1$.  The minimum
of~0.69 at $a_1\approx90^\circ$ represents a ${\sim}30\%$ cancellation,
requiring joint marginalisation over both systematics.}
\label{fig:degeneracy}
\end{figure}
\subsection{Interplay Between PREM and Planck-Scale Effects}

The two corrections enter the oscillation probability through distinct but
coupled channels (Eq.~(\ref{eq:Htotal})).  \prem-induced distortions are
primarily baseline-dependent, arising from energy-dependent phase shifts in the
appearance probability caused by the spatial variation of the MSW potential.
Planck-scale corrections shift the physical mass-squared differences and mixing
angles at the level of the underlying oscillation parameters, producing a
baseline-independent offset in the oscillation frequency.

At short baselines ($L<5000$\,km) the Planck-scale correction to $\Delta'_{21}$
of order $10^{-5}$\,eV$^{2}$ represents a potentially comparable systematic,
particularly for reactor-antineutrino experiments with high sensitivity to the
solar oscillation frequency.  At intermediate baselines (5000--7000\,km) the
\prem\ bias begins to dominate, but the Planck-scale correction acts as a
sub-leading, non-negligible systematic whose sign and magnitude depend on the
unknown Majorana phases.  At the longest baselines ($L>10000$\,km) the \prem\
effect is catastrophic and the Planck-scale correction is secondary but
non-negligible in $\chi^{2}$ fits that marginalise over both parameters
simultaneously.

\section{Future Directions and Conclusions}
\label{sec:conclusions}

\subsection{Future Directions}

\textbf{Full continuous \prem\ profile.}  The four-shell model is a significant
improvement over constant density but remains a discretisation.  Incorporating
the continuous \prem\ density profile---including transition zones between
shells---will reduce residual modelling bias at the few-percent level.

\textbf{Marginalisation over Earth model uncertainties.}  Future analyses
should propagate uncertainties in the \prem\ profile---arising from geophysical
measurement errors and regional deviations from spherical symmetry---as a
correlated systematic in the $\chi^{2}$ fit.  This is directly relevant to
atmospheric neutrino analyses with IceCube-Upgrade and KM3NeT/ORCA.

\textbf{Planck-scale corrections at current baselines.}  Although
present-generation LBL experiments (T2K, NOvA, \dune\ at $L\sim1300$\,km) are
unaffected by \prem\ stratification, the Planck-scale correction to
$\Delta'_{21}$ could produce observable effects in long-running high-statistics
reactor analyses, warranting dedicated study.

\textbf{Implications for \dune\ simulation software.}  Current \dune\ analyses
based on \globes\ employ a single effective matter density
($\rho\approx2.848$\,g\,cm$^{-3}$ for the 1285\,km baseline).  While adequate
at the \dune\ baseline, any extension of the \globes\ framework to simulate
longer-baseline proposals must replace this constant-density treatment with a
spatially resolved \prem\ profile~\cite{Pandit2026}.  Our results quantify the cost of not doing
so: a $17.8^{\circ}$ bias at $L=7000$\,km and a catastrophic $172.2^{\circ}$
bias at $L=12000$\,km.

\textbf{Detector resolution effects.}  The present analysis assumes perfect
energy resolution.  Incorporating realistic detector smearing would broaden
oscillation features and may raise the baseline threshold at which the
constant-density approximation becomes inadequate, but our preliminary analysis
(Section~\ref{sec:results}) shows that the bias exceeds the \dune\ Phase~II
target even with 20\% resolution.

\subsection{Conclusions}

We have presented the first unified analysis of Earth matter-density
stratification and Planck-scale quantum gravity corrections in long-baseline
neutrino oscillation experiments, with the novel result that these two
systematics are not independent but form a degenerate pair in $\chi^{2}$
analyses at intermediate baselines (5000--7000\,km).  Our principal conclusions
are:

\begin{enumerate}

\item The constant-density approximation introduces a bias
      $|\Delta\delta_{CP}|<0.3^{\circ}$ for $L<5000$\,km, but grows sharply
      to $17.8^{\circ}$\,(NH) / $14.2^{\circ}$\,(IH) at $L=7000$\,km and
      $172.2^{\circ}/163.7^{\circ}$ at $L=12000$\,km, the latter a complete
      sign reversal of reconstructed CP violation.

\item At $L=7000$\,km the bias exceeds the \dune\ Phase~II $1\sigma$ target of
      ${\sim}5^{\circ}$ by a factor of 3.6\,(NH), demonstrating that this is a
      fundamental systematic obstacle rather than a conservative refinement.

\item These results hold for both mass orderings; the IH bias at $L=7000$\,km
      ($14.2^{\circ}$) is nearly $3\times$ the \dune\ Phase~II target.

\item Planck-scale gravitational perturbations shift $\Delta m_{21}^{2}$ by
      $(1.0\pm0.5)\times10^{-5}$\,eV$^{2}$ for a degenerate neutrino mass
      spectrum, depending on the unknown Majorana phases.  The atmospheric
      splitting $\Delta'_{31}$ remains effectively unchanged.

\item The interplay between the two effects introduces a new class of degeneracy
      at $L\approx7000$\,km: at $a_{1}\approx90^{\circ}$ the combined bias is
      ${\sim}30\%$ smaller than the independent quadrature sum (degeneracy
      fraction~$=0.69$).  Joint marginalisation over both systematics is
      therefore essential.

\item These findings directly motivate the replacement of constant-density
      matter profiles in \globes-based \dune\ analyses with spatially resolved
      \prem\ implementations for any baseline beyond ${\sim}5000$\,km.

\end{enumerate}



\begin{thebibliography}{10}

\bibitem{Akhmedov2008}
E.~K. Akhmedov and V.~Niro.
\newblock An accurate analytic description of neutrino oscillations in matter.
\newblock {\em JHEP}, 12:012, 2008.

\bibitem{Abe2011}
K.~Abe et~al.
\newblock The {T2K} experiment.
\newblock {\em Nucl. Instrum. Meth. A}, 659:106--135, 2011.

\bibitem{Bian2013}
J.~Bian.
\newblock The {NOvA} experiment: overview and status.
\newblock {\em Nucl. Phys. B Proc. Suppl.}, 248--250:196--198, 2014.

\bibitem{DUNE2016}
R.~Acciarri et~al.
\newblock {Long-Baseline Neutrino Facility (LBNF) and Deep Underground Neutrino
  Experiment (DUNE): Conceptual Design Report, Volume~1}.
\newblock {\em arXiv e-prints}, arXiv:1512.06148, 2016.

\bibitem{Wolfenstein1978}
L.~Wolfenstein.
\newblock Neutrino oscillations in matter.
\newblock {\em Phys. Rev. D}, 17:2369--2374, 1978.

\bibitem{Smirnov2005}
A.~Yu. Smirnov.
\newblock The {MSW} effect and solar neutrinos.
\newblock {\em Phys. Scripta}, T121:57--64, 2005.

\bibitem{Pandit2026}
T.~Pandit and B.~S. Koranga.
\newblock Earth-density effects in {LBL} experiments: a comprehensive review
  of theory, observations, and future directions.
\newblock {\em arXiv e-prints}, arXiv:2601.21256, 2026.

\bibitem{Weinberg1979}
S.~Weinberg.
\newblock Baryon and lepton nonconserving processes.
\newblock {\em Phys. Rev. Lett.}, 43:1566--1570, 1979.

\bibitem{Vissani2003}
F.~Vissani, M.~Baldo, and G.~F. Burgio.
\newblock Revisiting the {Planck} scale effects on neutrino masses.
\newblock {\em Phys. Lett.\ B}, 571:209--214, 2003.

\bibitem{Koranga2008}
B.~S. Koranga, M.~Narayan, and S.~Uma Sankar.
\newblock Planck scale effects in neutrino oscillations.
\newblock {\em Phys. Lett.\ B}, 665:63--68, 2008.

\bibitem{Koranga2009}
B.~S. Koranga, M.~Narayan, and S.~Uma Sankar.
\newblock Planck scale corrections to neutrino mixing angles and {CP} violation.
\newblock {\em Fizika B}, 18:219--226, 2009.

\bibitem{Kelly2018}
K.~J. Kelly and S.~J. Parke.
\newblock Matter density profile shape effects at {DUNE}.
\newblock {\em Phys. Rev. D}, 98:015025, 2018.

\bibitem{Ohlsson2000}
T.~Ohlsson and H.~Snellman.
\newblock Three-flavour neutrino oscillations in matter.
\newblock {\em Phys. Rev. D}, 62:073004, 2000.

\bibitem{Giunti2007}
C.~Giunti and C.~W. Kim.
\newblock {\em Fundamentals of Neutrino Physics and Astrophysics}.
\newblock Oxford University Press, Oxford, 2007.

\bibitem{Dziewonski1981}
A.~M. Dziewonski and D.~L. Anderson.
\newblock Preliminary reference {Earth} model.
\newblock {\em Phys. Earth Planet. Inter.}, 25:297--356, 1981.

\end{thebibliography}
\end{document}